\documentclass[superscriptaddress, twocolumn, prb, showpacs]{revtex4}  
\usepackage{amssymb}
\usepackage{amsmath}
\usepackage{bbm}
\usepackage{graphicx}
\begin{document}

\author{Philipp Werner}
\affiliation{Department of Physics, University of Fribourg, 1700 Fribourg, Switzerland}
\author{Naoto Tsuji}
\affiliation{Department of Physics, University of Fribourg, 1700 Fribourg, Switzerland}
\author{Martin Eckstein}
\affiliation{Max Planck Research Department for Structural Dynamics, University of Hamburg-CFEL, Hamburg, Germany}

\title{Nonthermal symmetry broken states in the strongly interacting Hubbard model}

\date{\today}

\begin{abstract}
We study the time evolution of the antiferromagnetic order parameter after interaction quenches in the Hubbard model. Using the 
nonequilibrium dynamical mean field formalism, we show that the system, after a quench from intermediate to strong interaction, 
is trapped in a nonthermal state which is reminiscent of a photo-doped state and protected by the slow decay of doublons. If the 
effective doping of this state is low enough, it exhibits robust antiferromagnetic order, even if the system is highly excited and the 
thermal state thus expected to be paramagnetic. 
We comment on the implication of our findings for the stability of nonthermal superconducting states.      
\end{abstract}

\hyphenation{}


\maketitle

\noindent

\section{Introduction}

The rapid progress in the development of time-resolved spectroscopies enables direct measurements of dynamical symmetry breaking or 
the melting of long-range order in correlated materials.\cite{Cavalieri2007} Ultra-fast demagnetization after femto-second optical pulses 
was first observed in ferromagnetic Ni.\cite{Beaurepaire1996} The transient closing of the charge density wave 
(CDW) gap and excitations of CDW amplitude oscillations in 1T-TaS$_2$ 
\cite{Perfetti2006, Perfetti2008,Petersen2011,Hellmann2010,Eichberger2010} 
and TbTe$_3$ \cite{Schmidt2008} were extensively studied 
using time-resolved photoemission. These experiments showed that the CDW gap disappears within less than 100 femto-seconds 
(essentially the time-resolution of the experiment). Ref.~\onlinecite{Beaud2009} reported the melting of orbital order and related structural 
transitions in Manganites after a strong laser pulse. While the above studies focus on the destruction of some long-range ordered 
state (or at least the associated gap in the electronic spectrum) by a strong excitation, and emphasize the fast time-scale on which 
this process happens, there have also been recent experiments on photo-stimulated cuprates, which indicate the emergence of a 
nonthermal superconducting state which appears to be remarkably stable.\cite{Fausti2011}

Since all of these experiments involve transition metal compounds, they raise fundamental questions about the dynamics of symmetry 
breaking transitions and the stability of symmetry broken states in strongly interacting electron systems. While the concept of a phase 
transition is well understood in equilibrium, it is not obvious how to apply these ideas if a transition occurs under nonequilibrium conditions.
For example, it is unclear whether and how the timescale for the melting of (quasi) long-range order is related to 
equilibrium correlation times. A recent theoretical study of a dynamical Kosterlitz-Thouless transition found an evolution 
through nonthermal (superheated) states after a quench, and showed that the transition to a disordered state can become very slow if the latter is close to the equilibrium phase 
transition.\cite{Mathey2009} This example indicates that even a strongly excited nonequilibrium state can differ from a high-temperature disordered state on long timescales.

\section{Model and Method}

In this paper, we explore the dynamics of a symmetry broken state in the most fundamental model for correlated electron materials, the Hubbard model, 
after an interaction quench in the strongly correlated regime. The Hamiltonian is
\begin{equation}
H(t) = \sum_{ij,\sigma} V_{ij}c^\dagger_{i\sigma}c_{j\sigma} + U(t)\sum_i \left(n_{i\uparrow}-\tfrac12\right)\left(n_{i\downarrow}-\tfrac12\right),
\end{equation}
where $V_{ij}$ is the hopping amplitude between sites $i$ and $j$, $\sigma$ is the spin index, and the local interaction between electrons 
of opposite spin is $U$. We choose the hoppings such that the density of states becomes semi-elliptical, $\rho(\epsilon)=\sqrt{4V^2-\epsilon^2}/(2\pi V^2)$, 
and restrict our study to half-filling. Energy is measured in units of $V$ and time in units of $V^{-1}$. To solve this model, we employ the dynamical mean field 
theory (DMFT),\cite{Georges96} which has been extensively used to characterize the equilibrium phase diagram, and gives a qualitatively correct description 
for lattices with large coordination number.\cite{Metzner89} The DMFT formalism has recently been reformulated for nonequilibrium situations,\cite{Schmidt2002, Freericks2006} 
which gave insights into quench dynamics in the Hubbard model,\cite{Eckstein2009, Eckstein10quench} and to nonequilibrium states induced by external 
electric fields.\cite{Eckstein11pump, Eckstein2011, Eckstein11bloch, Tsuji2011, Tsuji2012, Aron2012}

\begin{figure}[t]
\begin{center}
\includegraphics[angle=-90, width=0.675\columnwidth]{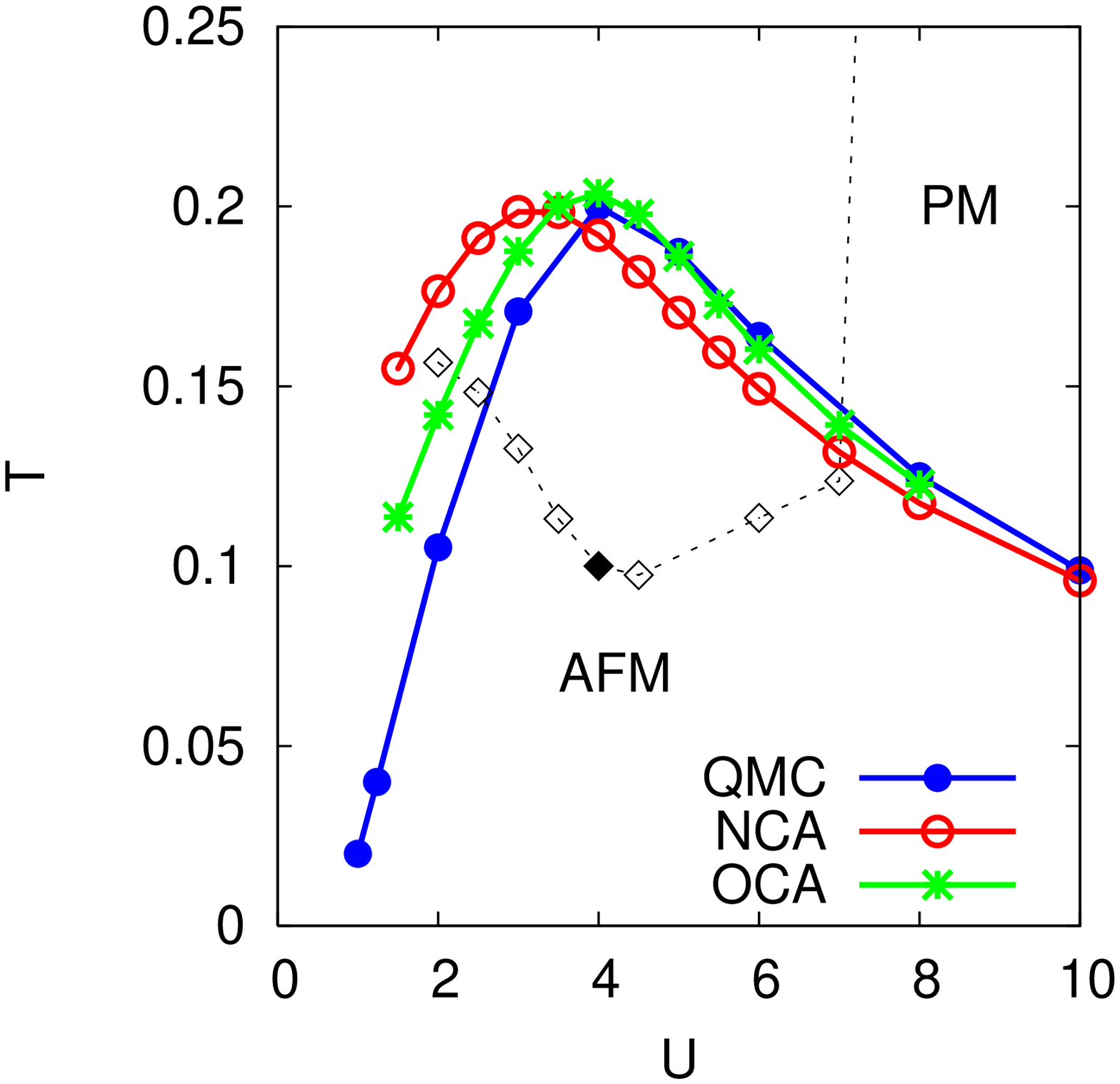}
\hfill
\includegraphics[angle=-90, width=0.29\columnwidth]{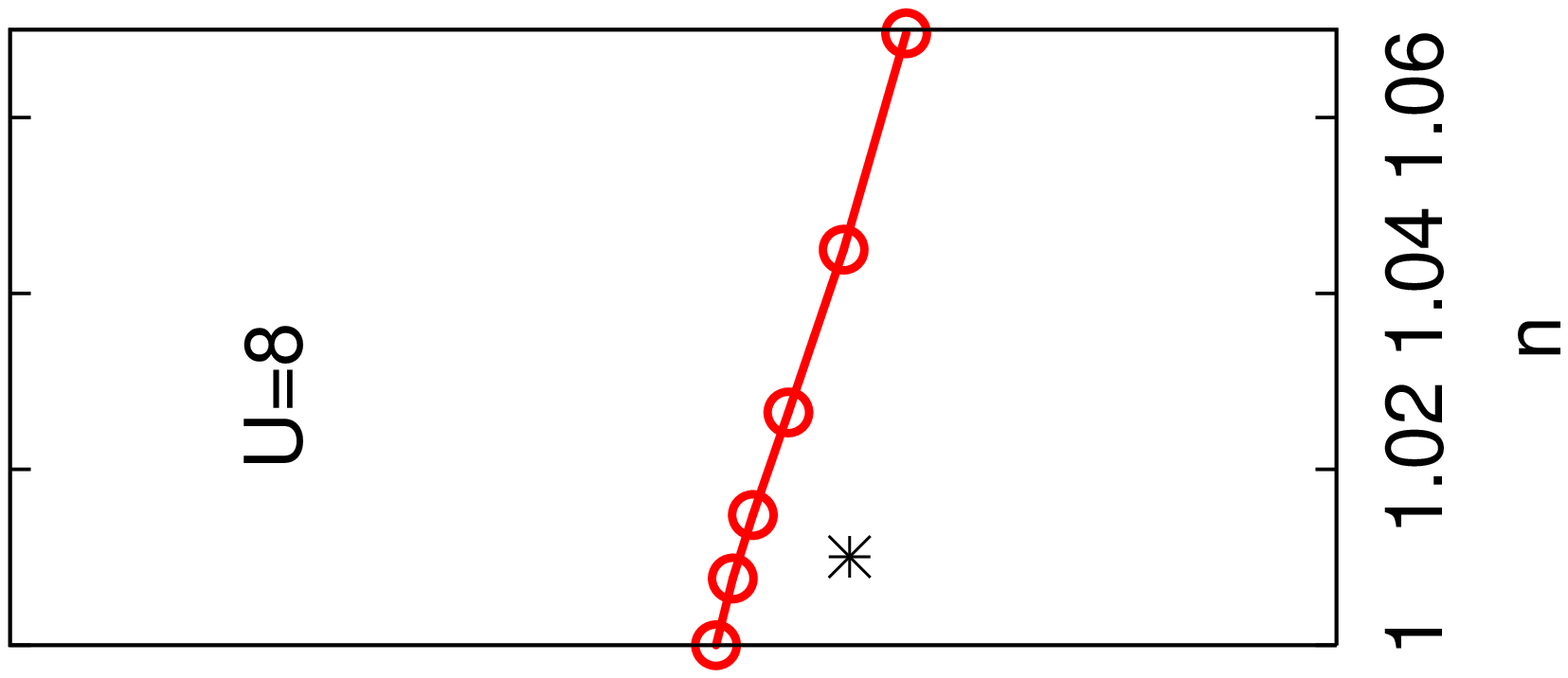}
\caption{
(Color online). Left panel: Antiferromagnetic phase diagram for the half-filled Hubbard model. The blue line with circles shows the exact DMFT result (from Ref.~\onlinecite{Koga2011}), the red line with open circles the NCA approximation and the green line with stars the OCA approximation. Black diamonds indicate $T_\text{eff}(U)$ for interaction quenches from $U=4$, $T=0.1$. Right panel: NCA phase boundary as a function of filling at $U=8$. The star indicates the effective temperature of the doped Hubbard model with the same magnetization as in the trapped state (see text).
}
\label{phasediagram}
\end{center}
\end{figure}

Nonequilibrium DMFT studies up to now have been restricted to the paramagnetic phases of the model,
although the extension of DMFT to symmetry-broken phases is straightforward.\cite{Georges96}
If the symmetry between spin-up and spin-down Green functions is not enforced, then the 
(single site) DMFT phasediagram for the half-filled, repulsive Hubbard model exhibits an 
antiferromagnetically ordered phase at low temperature (denoted by AFM in Fig.~\ref{phasediagram}). 
For attractive $U$, one finds an analogous phase diagram with AFM order replaced by $s$-wave 
superconductivity (at half-filling the superconducting state is degenerate with a CDW phase, but 
in the doped system, superconductivity is more stable). The nature of the $s$-wave superconducting 
(or AFM insulating) state changes qualitatively as $|U|$ crosses the value corresponding roughly to 
the maximum in the critical temperature. This is known as the ``BCS-BEC" crossover in the cold-atom 
literature. Here, we will focus on the strongly interacting ``BEC" (or localized moment / ``Mott") regime,
and study the stability of the ordered phase after a sudden switch of $U$ into the paramagnetic 
regime.  The quench provides an idealized but theoretically well-controlled excitation procedure.
Although it is artificial from the point of view of condensed matter experiments, we will see below that 
a qualitatively similar behavior of the long-time dynamics can be expected in photo-doping experiments.

\begin{figure}[t]
\begin{center}
\includegraphics[angle=-90, width=\columnwidth]{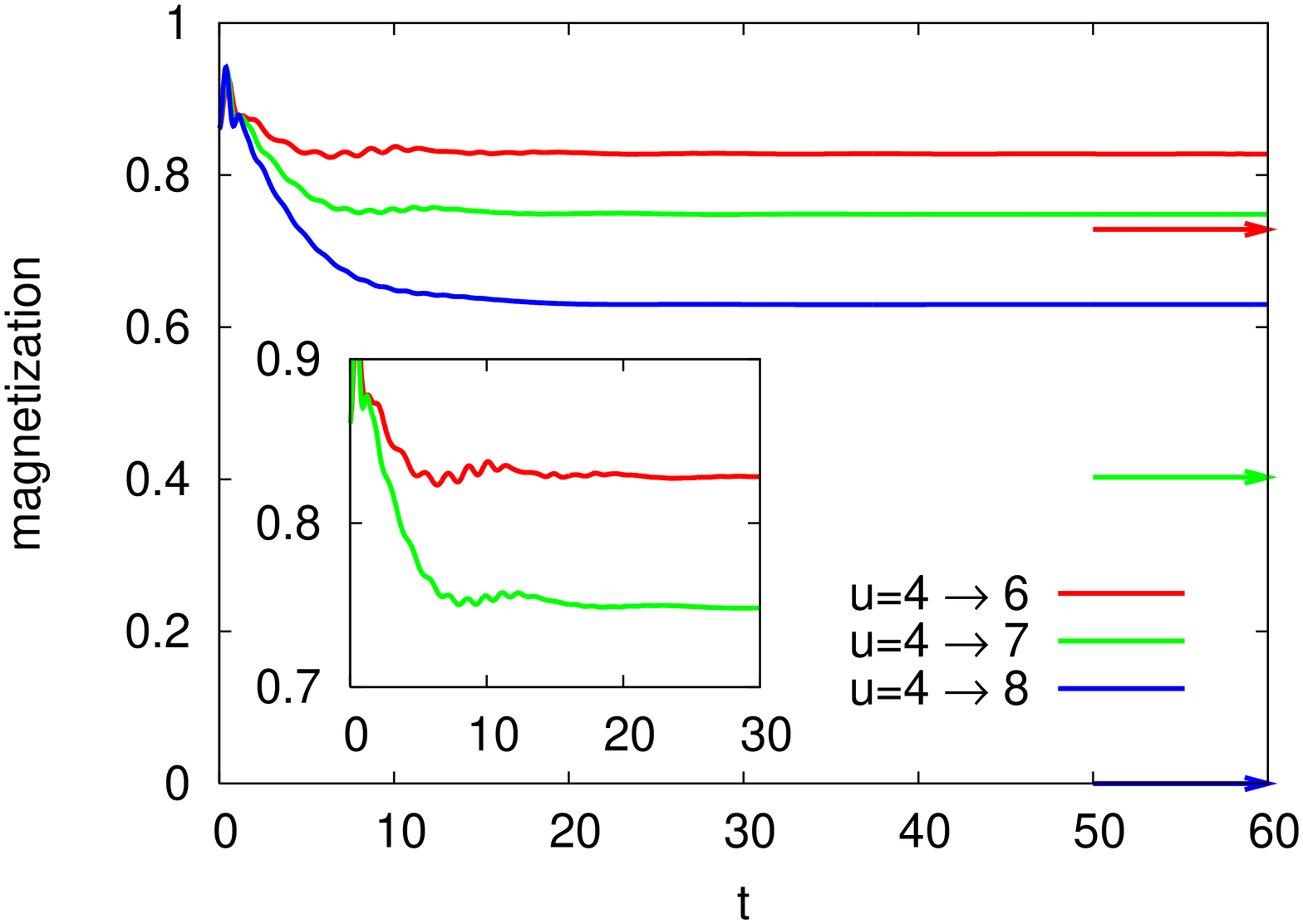}
\includegraphics[angle=-90, width=\columnwidth]{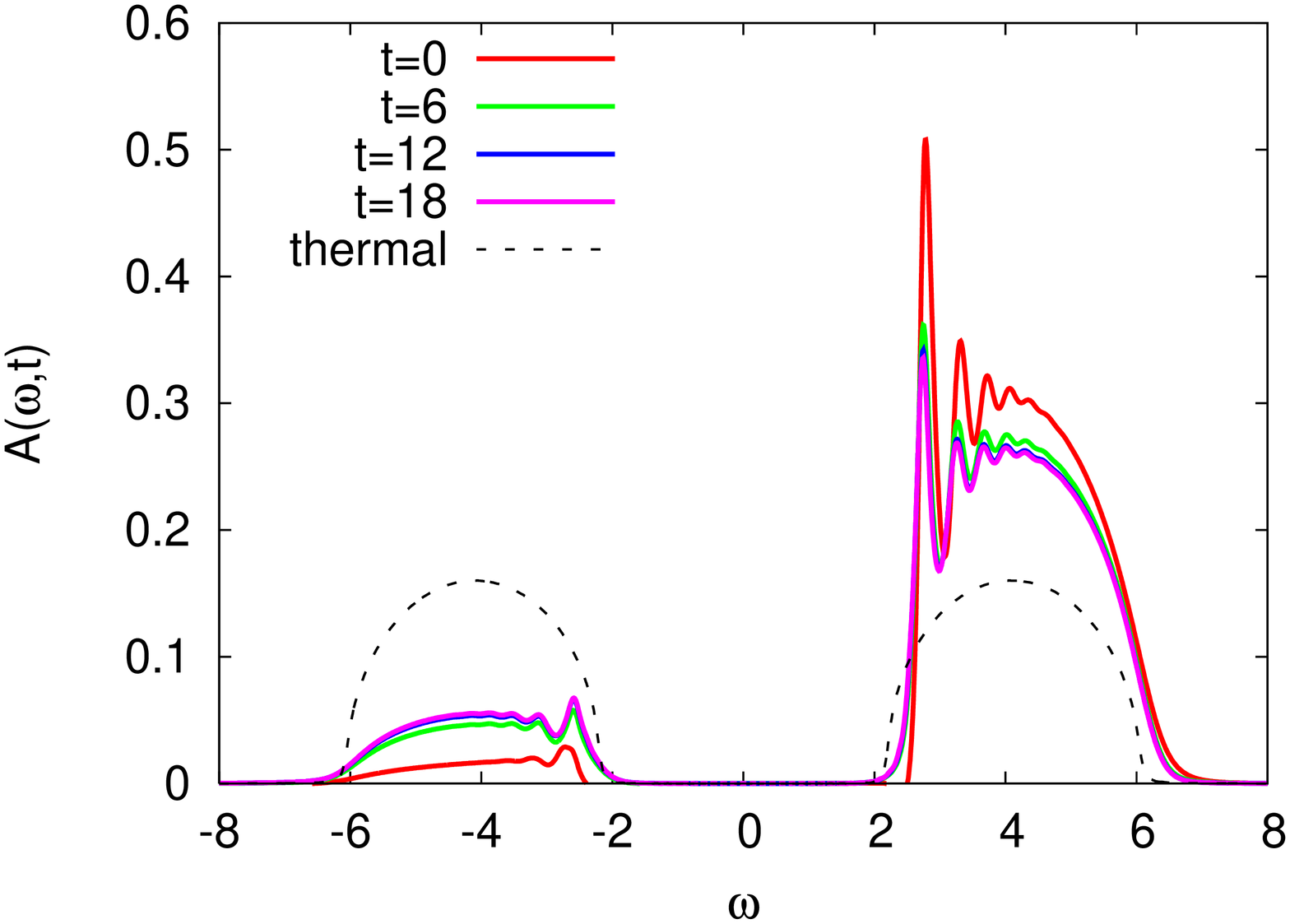}
\caption{(Color online). Top panel: Time evolution of the magnetization for quenches from $T=0.1$ and $U=4$ to $U=6$, $7$ and $8$. 
The effective temperatures after these quenches are $T_\text{eff}=0.113$, $0.124$ and $T=0.732$, respectively, and the arrows 
indicate the corresponding thermal values of the magnetization. Inset: same data on a different scale. Bottom panel: Time evolution of $A(\omega,t)$ for the minority spin after a quench from $T=0.1$ and $U=4$ to $U=8$, and comparison to the thermal result (dashed). The pink curve corresponds 
to the spectral function of the long-lived nonthermal state.}
\label{mag}
\label{spectrum}
\end{center}
\end{figure}

To study antiferromagnetic states within DMFT on a bipartite lattice, we have to solve impurity problems for each sublattice. For the semi-elliptic 
density of states (Bethe lattice), the hybridization function $\Lambda_{A,\sigma}$ ($\Lambda_{B,\sigma}$) for the $A$ ($B$) sublattice is given
by the self-consistency condition $\Lambda_{A,\sigma}=V^2G_{B,\sigma}$ ($\Lambda_{B,\sigma}=V^2G_{A,\sigma}$), where $G$ is the local
lattice Green function. Together with the relation $\Lambda_{A,\sigma}=\Lambda_{B,\bar\sigma}$ (for pure Neel-type symmetry breaking), this leads
to a single impurity calculation with self-consistency $\Lambda_\sigma=V^2G_{\bar\sigma}$.
To solve the nonequilibrium DMFT impurity problem, we use self-consistent strong-coupling perturbation theory. Figure~\ref{phasediagram} 
shows the equilibrium phase diagram obtained with the two lowest order implementations: the non-crossing approximation (NCA) 
\cite{Keiter1971} and the one-crossing approximation (OCA).\cite{Pruschke1989} While NCA provides a rather good description of 
the exact QMC phase boundary \cite{Koga2011} in the strongly correlated regime, the maximum is shifted to 
$U_\text{max}^\text{NCA}\approx 3$ (while $U_\text{max}^\text{QMC}\approx 4$) and the phase boundary in the weakly correlated 
regime is not well reproduced. OCA brings a significant improvement, with an almost correct position of the maximum and a quantitatively 
accurate description of the phase boundary in the strong correlation regime. Since we are mainly interested in quenches
within the strongly correlated regime, we will use the NCA 
method for the real-time calculations. The techniques for the solution of the nonequilibrium DMFT equations and our implementation of the 
real-time NCA/OCA impurity solver have been explained in detail in 
Refs.~\onlinecite{Eckstein10quench} and  \onlinecite{Eckstein10nca}.        

We measure the time-dependent expectation values of the magnetization $m=n_\uparrow-n_\downarrow$, the double occupancy 
$d=n_\uparrow n_\downarrow$ (which yields the local energy $E_\text{loc}=Ud-\mu(n_\uparrow+n_\downarrow)$), and the kinetic 
energy per spin $E_{\text{kin},\sigma} = -\frac{i}{L} \sum_{ij}   V_{ij} G_{ji,\sigma}^<(t,t)$ ($G_{ji,\sigma}^<(t, t')=i\langle c^\dagger_{i,\sigma}(t')c_{j,\sigma}(t) \rangle$). 
The latter can be expressed (within DMFT) as a convolution of the local Green function $G_{ii,\sigma}$ and the hybridization function 
$\Lambda_{i,\sigma}$. With $\Sigma_{i,\sigma}$ denoting the self-energy for site $i$, we can write the lattice and impurity Dyson equations as 
$[ i\partial_t + \mu - \Sigma_{i,\sigma}] G_{ij,\sigma} - \sum_k V_{ik} G_{kj,\sigma} = \delta_{ij}$ and $[ i\partial_t + \mu - \Sigma_{i,\sigma} - \Lambda_{i,\sigma} ] G_{ii,\sigma} = 1$. 
Hence, $\sum_k V_{ik} G_{ki,\sigma} = \Lambda_{i,\sigma} G_{ii,\sigma}$, and with sublattice indices $A$ and $B$ one obtains 
\begin{equation}
E_{\text{kin},\sigma} = -(i/2) [\Lambda_{A,\sigma}G_{A,\sigma} + \Lambda_{B,\sigma}G_{B,\sigma}  ]^<(t,t).
\end{equation}

\section{Results}  
  
\subsection{Magnetization and spectral function}  
  
The top panel of Fig.~\ref{mag} plots the time-evolution of the magnetization for quenches from an 
initial equilibrium state at $U=4$ and $T=0.1$, which is located deep inside the antiferromagnetic phase, to final states with $U=6$, $7$,  $8$. The
sudden increase of $U$ causes a change in the total energy $E_\text{tot}=E_\text{loc}+\sum_\sigma E_{\text{kin},\sigma}$ of the system. $E_\text{tot}$ stays constant after the 
quench and allows us to define an effective temperature $T_\text{eff}(U)$ corresponding to the temperature of the thermal state with interaction $U$ and 
total energy $E_\text{tot}$. This effective temperature is indicated by the black diamonds in Fig.~\ref{phasediagram}. 
While $T_\text{eff}$ initially decreases with increasing $U$ due to the decreasing slope of constant entropy curves in this part of the phase diagram, 
it increases above $U\approx 4.5$ and crosses the AFM phase boundary slightly above $U=7$. After the quench to $U=8$, the system is expected 
to thermalize in a paramagnetic (PM) state at high temperature ($T_\text{eff}(U=8)=0.732$).  However, as shown in Fig.~\ref{mag}, 
after a modest decrease the magnetization 
remains stuck for $t\gtrsim 20$ at some large, nonthermal value, with no further relaxation evident on the timescales accessible 
in our simulations. 
Similarly, the quenches to $U=6$ and $7$ lead to a trapping in a nonthermal state, as is evident from the comparison to the thermal value of the magnetization (arrows). While the transient dynamics also shows interesting behavior, in particular slow oscillations in the amplitude of the order parameter (with period $\approx 10$), with superimposed rapid $1/U$-modulations (see inset of Fig.~\ref{mag}), we will focus in this paper on the trapping phenomenon and the nature of the long-lived nonthermal state. 

To further characterize this state, we compute a time-resolved ``spectral function" $A(\omega,t)$ from the Fourier 
transform of the retarded Green function 
$G^\text{ret}(t,t')=-i\Theta(t-t')\langle\{c(t),c^\dagger(t')\}\rangle$: 
\begin{equation}
A(\omega,t)=-\frac{1}{\pi}\text{Im}\int_t^\infty dt' e^{i\omega(t'-t)}G^\text{ret}(t',t).
\end{equation}
This function is 
plotted for the minority spin in Fig.~\ref{spectrum}, for the quench from $U=4$, $T=0.1$ to $U=8$. While 
the red curve ($t=0$) should not be confused with the spectral function of the initial equilibrium state, it exhibits 
Hubbard bands
with pronounced spin-polaron peaks, as is typical for the AFM insulator.\cite{Taranto2012} 
Consistent with the left panel of Fig.~\ref{mag}, the spin imbalance shrinks from $t=0$ to $t\approx 18$ 
and then becomes time-independent for larger $t$. 
The spectral function of the trapped non-thermal state still features spin-polaron peaks, 
in contrast to 
the thermal spectral function at $T_\text{eff}=0.732$ (dashed lines). 
We conclude that despite the strong excitation of the system and the large amount of energy injected by the quench from 
$U=4$ to $U=8$, the antiferromagnetic order does {\it not} melt rapildy - instead, the system is trapped in a long-lived state 
with large magnetization, and with the typical spectral features of a magnetically ordered state. 

\begin{figure}[t]
\begin{center}
\includegraphics[angle=0, width=0.494\columnwidth]{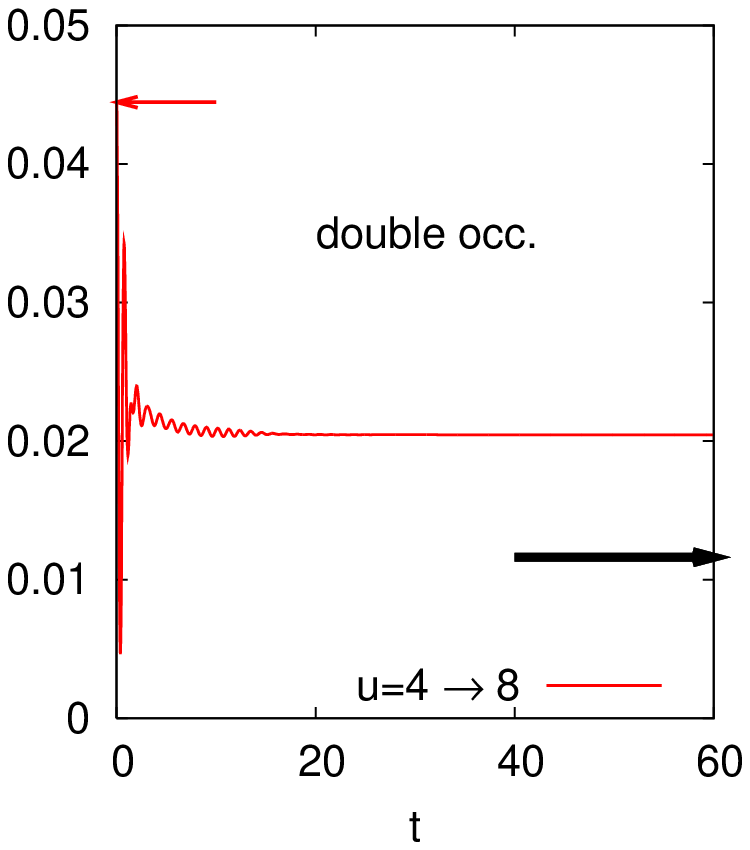}
\includegraphics[angle=0, width=0.494\columnwidth]{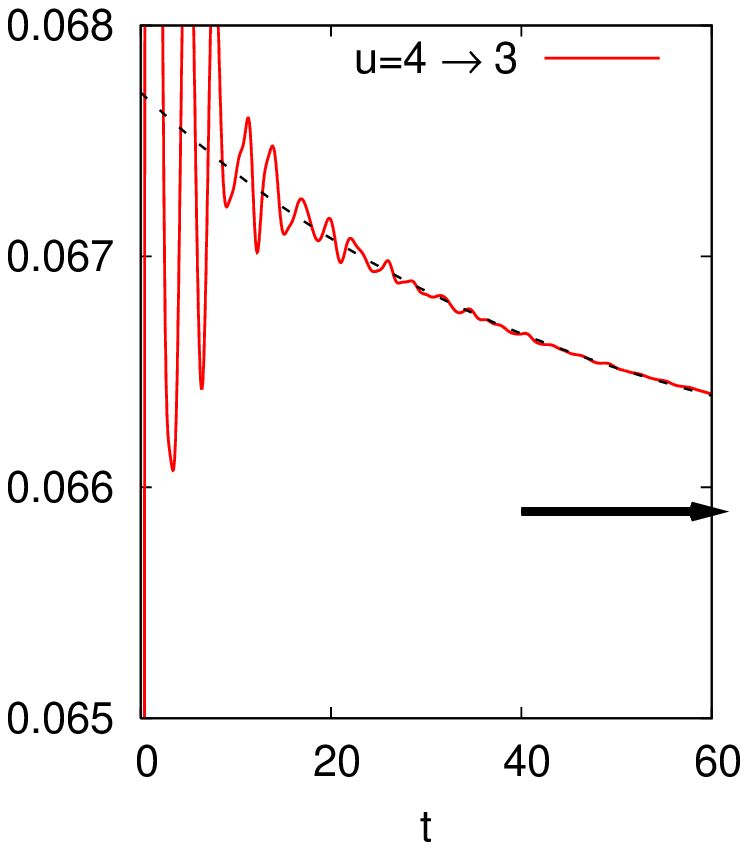}
\includegraphics[angle=0, width=0.494\columnwidth]{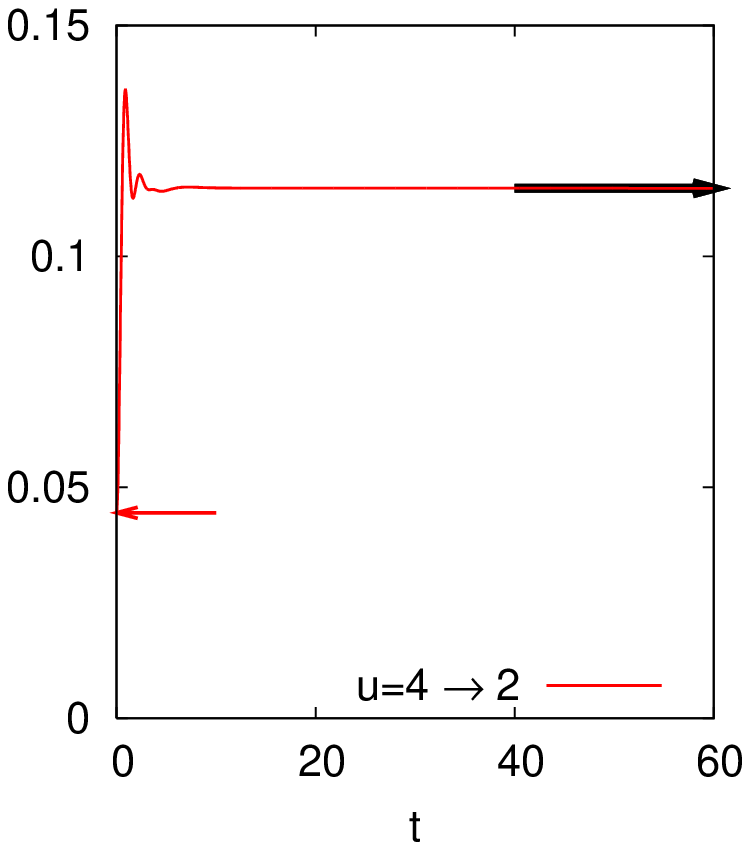}
\includegraphics[angle=0, width=0.494\columnwidth]{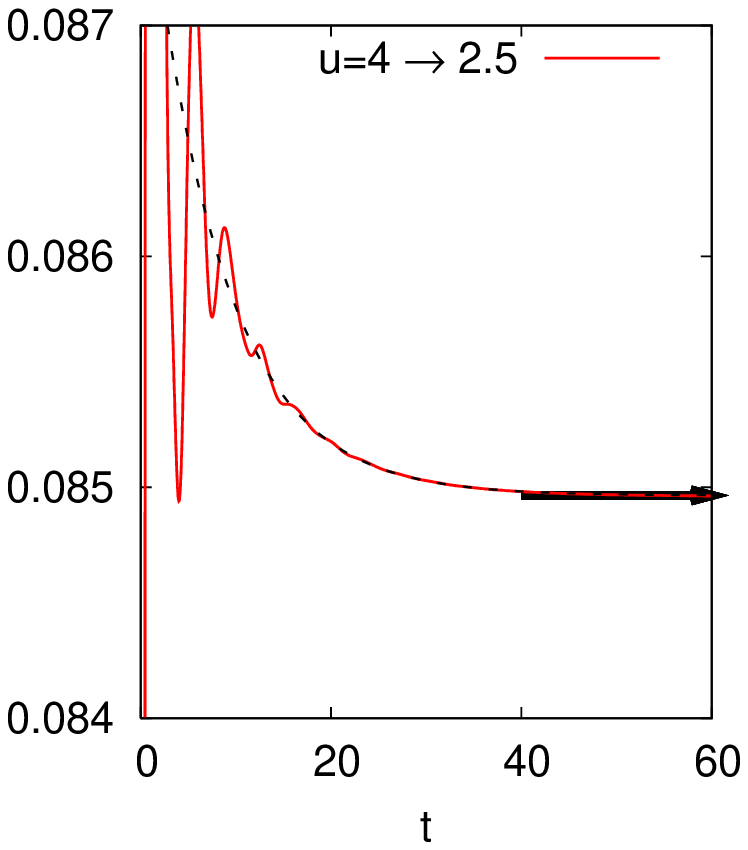}
\caption{
(Color online). Time evolution of the double occupancy for the quench from $U=4$, $T=0.1$ to $U=8$, $3$, $2.5$, $2$ 
(clockwise from the top left) and comparison to the thermal result (black arrow). The right panels show a narrow window around 
the thermal value and (as dashed line) an exponential fit to the long-time behavior. The left-pointing arrows in the left panels show the inital value of the double occupancy. 
}
\label{double}
\end{center}
\end{figure}

\subsection{Double occupancy}

To understand the nature of this trapped state and the reason for its robustness, we consider the time-evolution of the average number 
of doubly occupied sites. The top left panel of Fig.~\ref{double} shows the result for the quench from $U=4$, $T=0.1$ to $U=8$. 
There is a complicated transient regime (with $1/U$ oscillations) up to $t\approx 20$ in which the double occupancy decreases from 
about $0.044$ to $0.021$.
 At longer times, the double occupancy is stuck at $0.021$, even though the expected thermal value 
is $0.0115$ and thus almost a factor of two lower. 
This finding is consistent with theoretical and experimental studies of cold atom systems,\cite{Sensarma2010} and with 
a previous investigation of pump-excited paramagnetic Mott insulators \cite{Eckstein11pump}
which show that
the relaxation 
time of doublons grows exponentially with $U$, leading - at large $U$ - to a trapping in a nonthermal state characterized by a fixed 
number of doublons. 

Indeed, if one quenches to smaller interaction, the relaxation time decreases and eventually becomes measurable on the 
timescales accessible in the simulations (right hand panels of Fig.~\ref{double}). We have fitted the long-time behavior of the double 
occupancy with an  
exponential function decaying 
onto the thermal value and thus extracted the relaxation times $\tau\approx 400$ 
for quenches from $T=0.1$, $U=4$ to $U=3.5$, $\tau\approx 47$ ($U=3$) and $\tau\approx 7.8$ ($U=2.5$).
The $U$-dendence of the relaxation time   
agrees with the analytical formula\cite{Sensarma2010} $\tau=Ae^{\alpha (U/2)\log(U/2)}$, with $A=0.165$ and $\alpha=5.6$. 
For the paramagnetic phase studied in Ref.~\onlinecite{Eckstein11pump}, a smaller coefficient $\alpha$ was found, but due 
to the tails in the Gaussian DOS used there, the results cannot be directly compared by rescaling the hopping.

Around $U=2$ (close to the maximum critical temperature in NCA), the relaxation becomes so fast that the system already 
thermalizes within the time of the initial transient (bottom left panel of Fig.~\ref{double}). This finding of a qualitatively different 
relaxation pathway and fast thermalization at intermediate coupling is consistent with the results of Ref.~\onlinecite{Eckstein11pump} 
for pump-excitations of the paramagnetic Mott insulator, and also with the result of Ref.~\onlinecite{Eckstein2009} for interaction 
quenches from a noninteracting initial state. 

\subsection{Comparison to a doped state}

The trapping of the double occupancy is well understood in the strong-coupling limit: For $U\gg V$, a 
unitary Schrieffer-Wolff transformation $e^S$ can be constructed order-by-order in $V/U$, such that 
successively terms of all orders in $V/U$ are removed from the commutator $[e^{-S}He^{S},d]$, and 
hence $\bar d=e^{S} d e^{-S}$ is conserved on exponentially long times.\cite{MacDonald1988}
At second order, and after projection to $\bar d=0$, 
one would obtain the $t$-$J$ model, but here we encounter a more general situation, with $\bar d >0$.
One can consider $\bar d$ as a rigorous definition of the number of free doublons, which differs from $d$ 
by quantum fluctuations $[S,d]+...=\mathcal{O}(V/U)$ 
(even in the Mott insulator at $T=0$, where $\bar d=0$). 
The initial very fast drop of $d(t)$ on the timescale of a few inverse hoppings (Fig.~\ref{double}), 
is thus related to the reduction of quantum fluctuations due to the increase of $U$, or equivalently, a strengthening 
of the local moments. 
Because the quench is faster than the timescale of a quantum fluctuation, quantum fluctuations are transformed into 
real doublon and hole excitations with a certain 
amplitude. This
gives rise to a nonzero $\bar d$ for $t>0$, whose 
stability on exponentially long 
times prevents the system from thermalization. 
The argument 
suggests that both the properties of the 
trapped magnetic states and the melting of the antiferromagnetic order can be related to the presence of 
injected free doublons and holes. To further support 
this fact 
it is useful to take a closer look at the spectral function.

\begin{figure}[t]
\begin{center}
\includegraphics[angle=-90, width=\columnwidth]{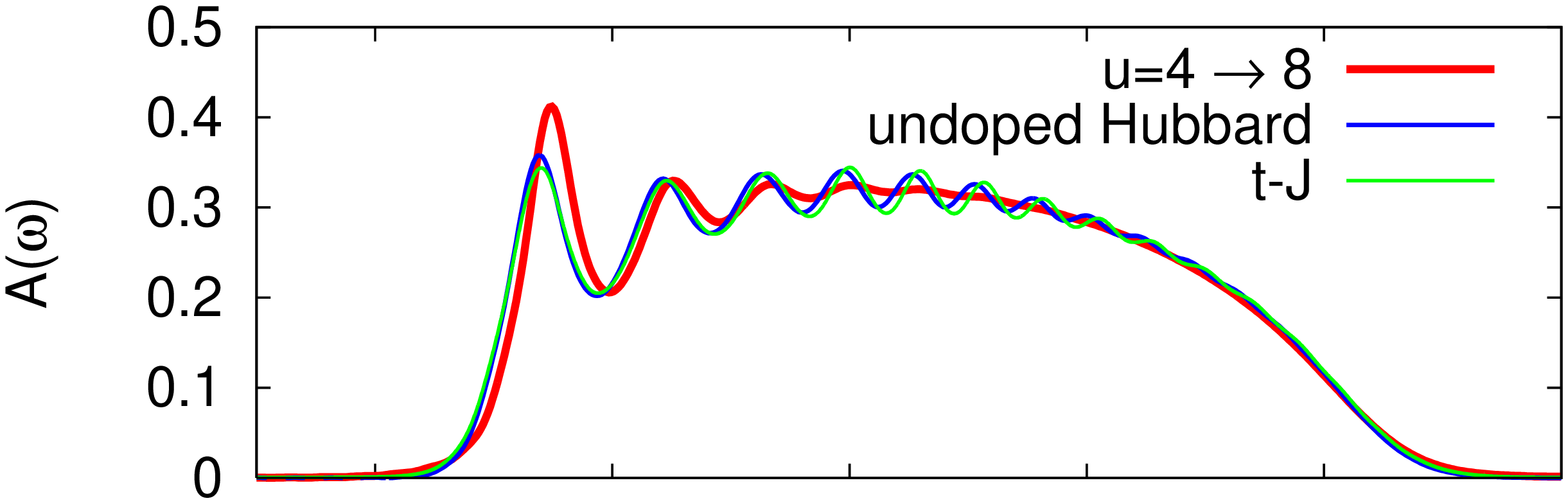}
\includegraphics[angle=-90, width=\columnwidth]{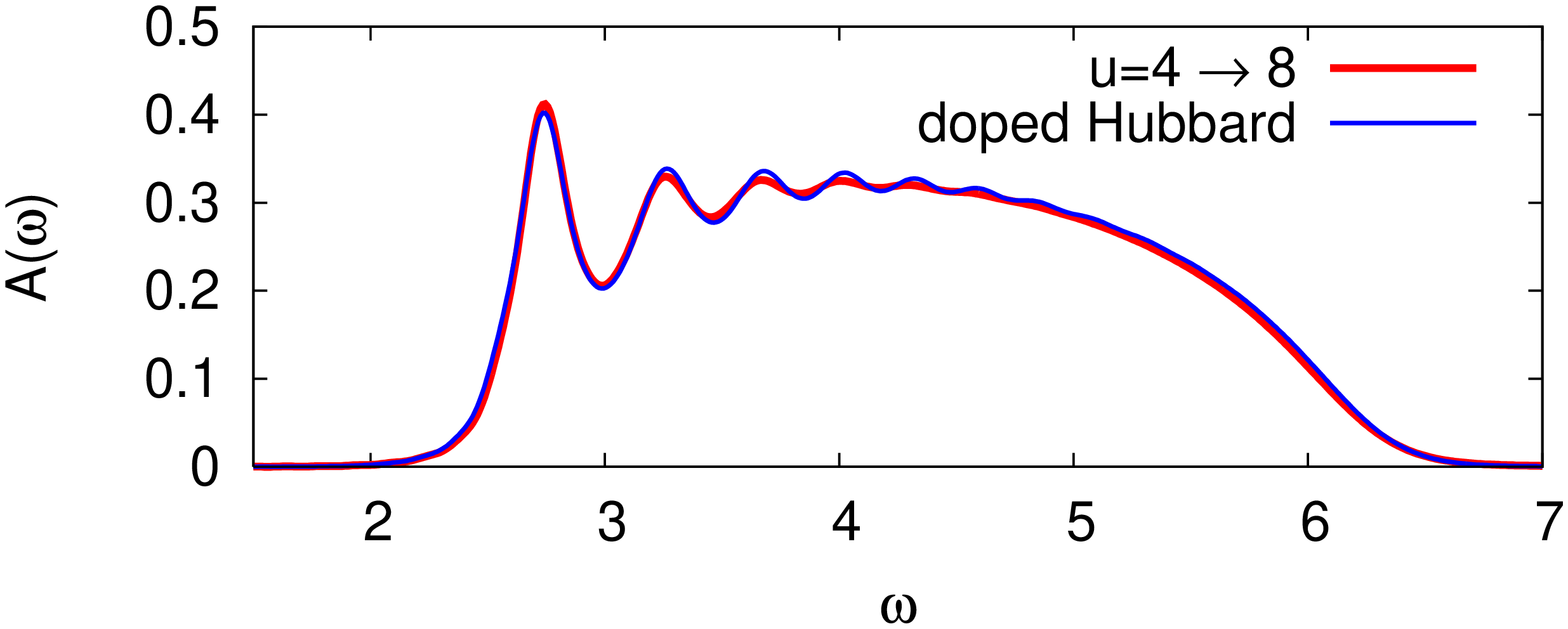}
\caption{
(Color online). Comparison of the spectral function $A(\omega,t=18)$ of the trapped state after the quench from $U=4$, $T=0.1$ 
to $U=8$ (bold red line) to the majority-spin spectral function of a half-filled Hubbard model with $T=0.0986$ and to an infintesimally 
doped  $t$-$J$ model with $J=4t^2/U=0.5$ (top panel). The blue curve in the bottom panel corresponds to a doped 
Hubbard model with $T=0.0917$ and chemical potential chosen such that the magnetization and double-occupancy of the 
trapped state are reproduced. (The $t$-$J$ and doped Hubbard spectra are shifted by $+4.2$ and $+2.4$ respectively.)  
}
\label{tj}
\end{center}
\end{figure}

In Fig.~\ref{tj} we compare the spectral function of the trapped state ($t=18$) to equilibrium spectral functions 
of an infinitesimally doped $t$-$J$ model with $J=4t^2/U$ and identical magnetization, as well as to a doped 
AFM Mott insulator. The spectral function for one hole in the $t$-$J$ model 
can be computed exactly within DMFT.\cite{Strack1992, Logan1998} 
The top panel of Fig.~\ref{tj} reveals a good agreement of this $t$-$J$ result and the spectral function of the 
half-filled Hubbard model at $U=8$, which shows that $U=8$ is in the $t$-$J$ limit, and that the NCA solution 
of the Hubbard model in this interaction regime is reliable. (At non-zero temperature, the spectral functions of the half-filled and infinitesimally doped Hubbard model are identical.) The temperature
$T=0.0986$ of the half-filled Hubbard model has been adjusted such that the magnetization corresponds 
to that of the trapped state, while the $t$-$J$ spectrum has been rescaled by $(1+m)/2=0.815$, and the $t$-$J$ spectrum has been shifted on the frequency axis by $+4.2$. 

The figure also shows that the equilibrium results do not very well reproduce the peak structures in the spectral function 
of the trapped state.  
The first peak near the gap edge is more prominent, while the higher energy peaks are smeared out. 
According to the above strong-coupling argument, the trapped state might be close to an 
equilibrium state with additional free doublons and holes. 
Such a state with doublons and holes is not accessible in thermal equilibrium within the 
Hubbard model. In the paramagnetic case and  at low energies, its properties are different from
a chemically doped state at the same doping.\cite{Eckstein2012} Nevertheless, because 
the interaction of doublons with other doublons is very different from the interaction of doublons with holes, 
one may assume that the main correction to the spectral 
function at positive energies (i.e., for inserting a doublon) is due to the presence of other free doublons, which motivates a 
comparison of the  spectral function of the trapped state to an electron-doped equilibrium state.
We determined the doping level ($n=1.010$) and temperature ($T=0.0917$) of the equilibrium Hubbard model 
such that the magnetization and double occupancy of the trapped state is reproduced. The corresponding point in the 
$T$-versus-filling phasediagram (right hand panel of Fig.~\ref{phasediagram}) is indicated by the black star symbol. The 
bottom panel of Fig.~\ref{tj} shows that the spectral function of this doped Hubbard model 
(shifted on the frequency axis by $+2.4$) is much closer to that of the 
trapped state. In particular, the dominant first peak is well reproduced, while the damping of the peaks at higher 
energies is still not strong enough. This may be due to the missing interaction with holons.

\begin{figure}[t]
\begin{center}
\includegraphics[angle=-90, width=\columnwidth]{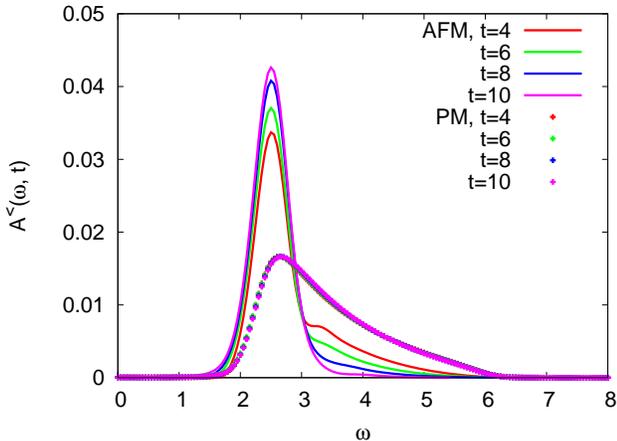}
\caption{Time-evolution of $A^<(\omega,t)$ for a quench from $U=4$, $T=0.1$ to $U=8$. 
While the occupied states rapidly accumulate at the lower band edge in the AFM calculation, the occupation is nearly time independent in the PM case. 
The curves are slightly broadened by introducing a Gaussian factor $e^{-\alpha(t-t')^2}$ in the integrand of Eq.~(\ref{spect_Alesser}), which ensures a smooth cutoff 
of the upper integration limit.
}
\label{Alesser}
\end{center}
\end{figure}

\subsection{Occupation function}

Our results suggest that the trapped state is essentially an AFM Mott insulator with simultaneous electron and hole doping, comparable to a state 
that can be prepared by ``photo-doping".
The doping level is determined 
by the density of trapped doublons and holons and the effective temperature is determined by the trapped magnetization. 
In particular, the effective temperature of the trapped state after a quench from $U=4$, $T=0.1$ to $U=8$ is 
apparently below the Neel temperature $T_N^\text{doped}$ of this photo-doped state. This remarkable fact must be the result of 
``entropy cooling": the reduction in the magnetization from $0.85$ to $0.63$ releases a lot of entropy and this in turn keeps the 
trapped state below its Neel temperature. 
Evidence for the cooling effect is provided by the occupied density of states,
\begin{equation}
A^<(\omega,t)=-\frac{1}{\pi}\text{Im}\int_t^\infty dt' e^{i\omega(t'-t)} \langle c^\dagger(t')c(t)\rangle,
\label{spect_Alesser}
\end{equation}
which corresponds to a time-resolved photoemission spectrum for a quasi-steady state (see Fig.~\ref{Alesser}). Since doublons and holes 
are inserted locally, they initially have a large kinetic energy, which leads to a broad distribution of 
the weight of $A^<(\omega,t)$ in the upper Hubbard band. While the magnetization is decreasing, the 
weight becomes concentrated at low frequencies (near the lower band edge), indicating that doublons and holes are being cooled 
by exchanging energy with the spin background. This is in contrast to a quench in the paramagnetic phase, 
after which the occupation is not redistributed in 
time,\cite{Eckstein11pump} 
because within DMFT the spin background of the 
paramagnetic state is completely disordered (no short range correlations) and hence does not allow 
energy exchange with doublons and holes.

\begin{figure}[t]
\begin{center}
\includegraphics[angle=-90, width=0.494\columnwidth]{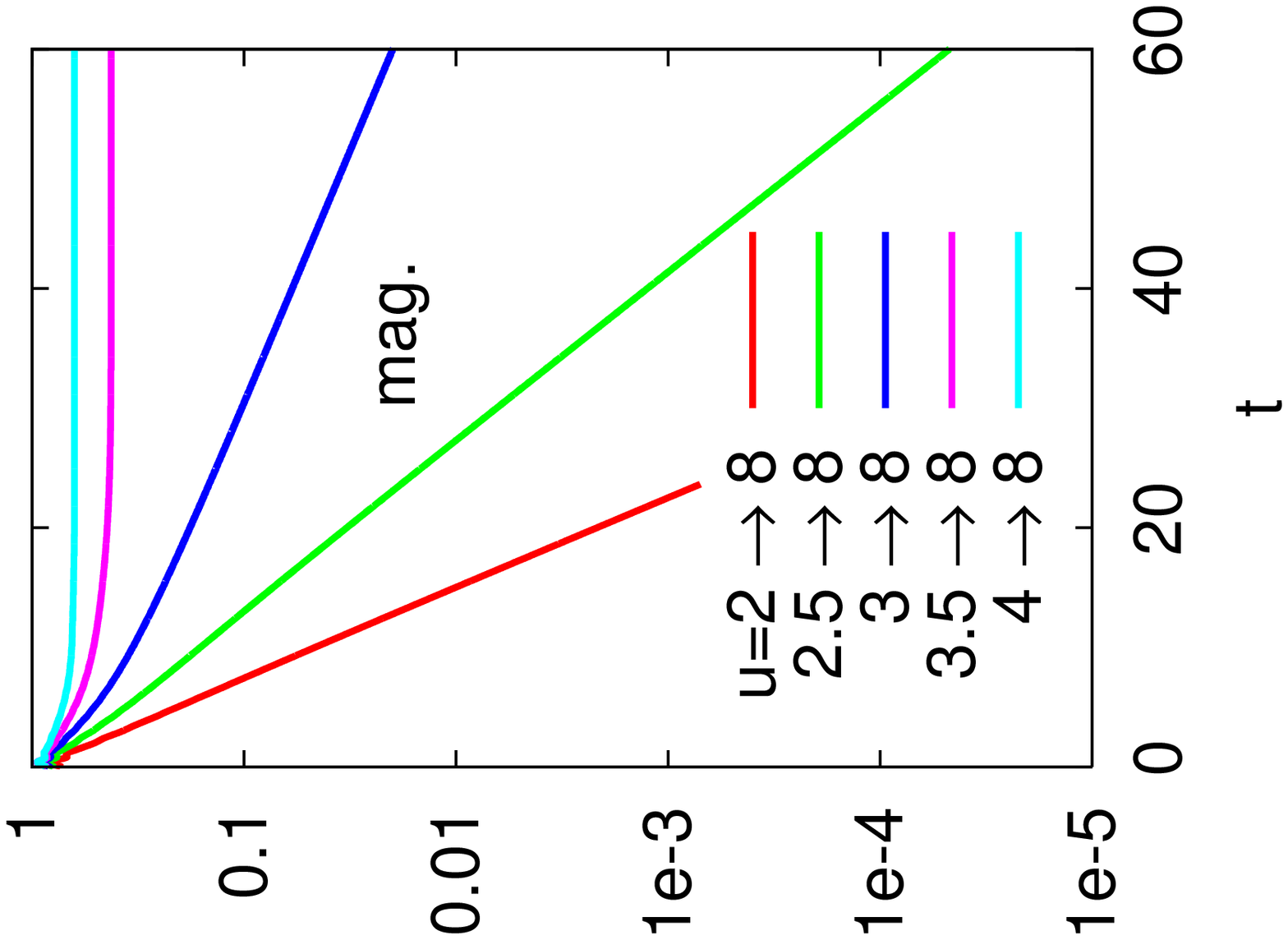}
\includegraphics[angle=-90, width=0.494\columnwidth]{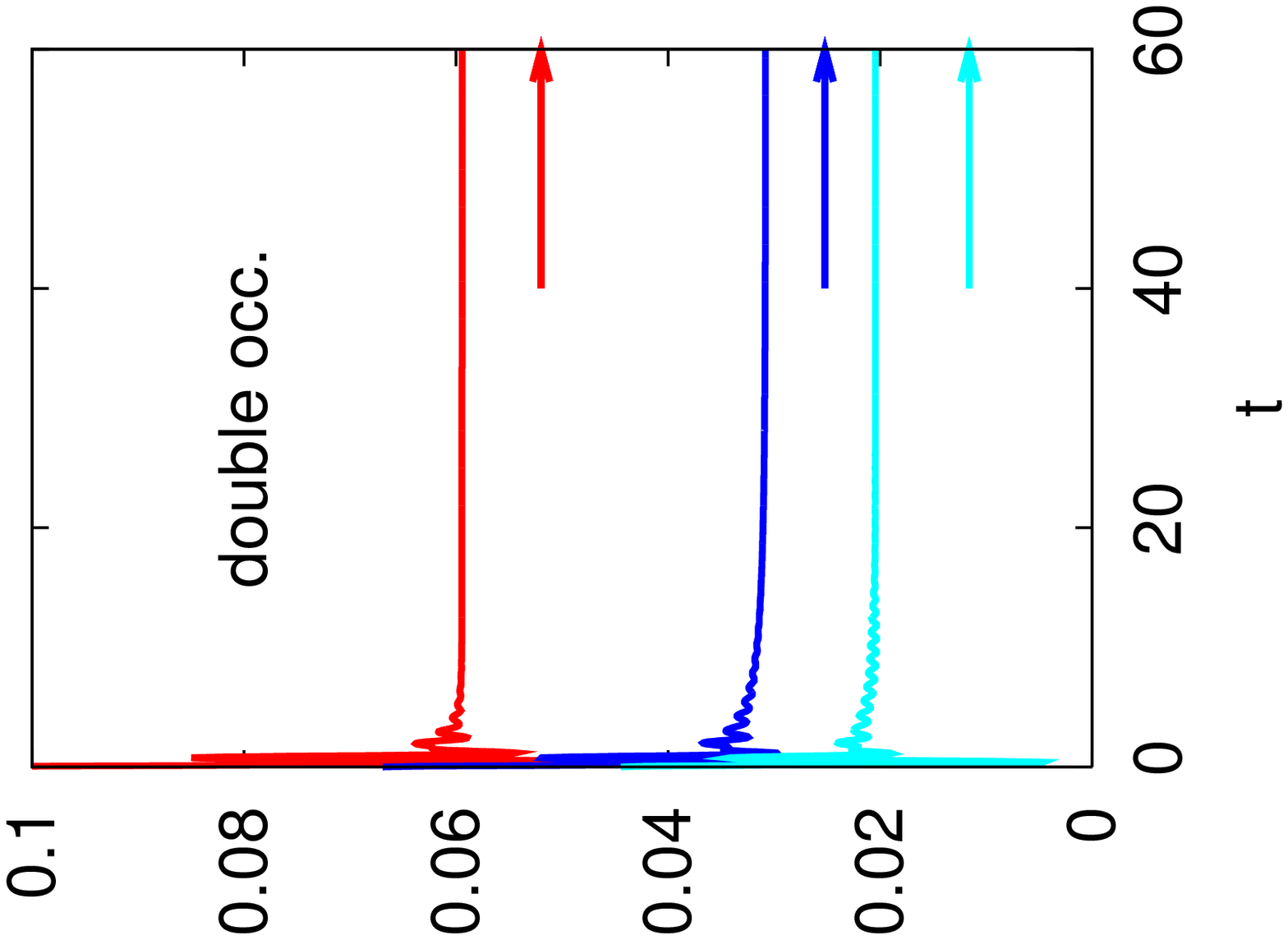}
\includegraphics[angle=-90, width=\columnwidth]{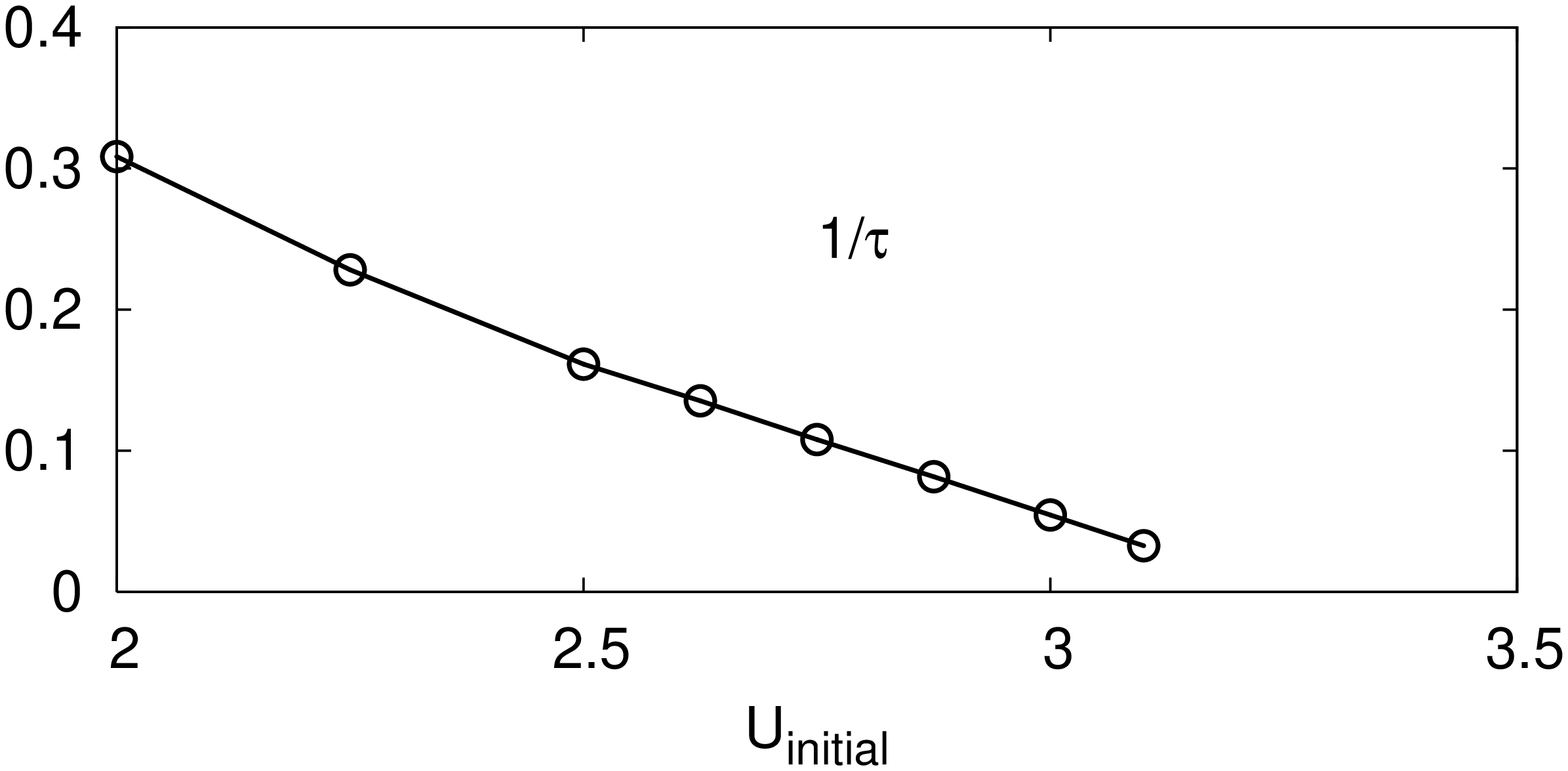}
\caption{Time-evolution of the magnetization (top left panel) and double occupancy (top right panel) for quenches from indicated initial $U$, $T=0.1$ to $U=8$. The bottom panel shows the inverse of the relaxation time $\tau$ as a function of initial $U$.
}
\label{doping}
\end{center}
\end{figure}

If the interpretation of entropy cooling 
is correct, one should be able to observe the transition of the trapped 
state from below $T_N^\text{doped}$ to above $T_N^\text{doped}$ in the nonequilibrium dynamics, by changing the initial density 
of doublons and holes ($T_N^\text{doped}$ decreases with doping, see right hand panel of Fig.~\ref{phasediagram}). Indeed, as 
illustrated in Fig.~\ref{doping}, the magnetization decays to zero exponentially for quenches from $U_\text{initial}\lesssim 3.3$, $T=0.1$ to $U=8$, 
while the system is trapped in magnetized states for quenches from $U_\text{initial}\gtrsim 3.3$, $T=0.1$. A smaller initial $U$ means a larger 
number of doublons and holes in the initial state, and thus also a larger doping of the effective photo-doped state. In other words, 
around $U_\text{initial}=3.3$, the effective photo-doped state crosses $T_N$. But even for $U_\text{initial}<3.3$, where the AFM 
order decays to zero after the quench, the system is still trapped for very long times in a nonthermal state, as can be seen for example by 
comparing the double occupancy in the trapped state to the thermal value (top right panel of Fig.~\ref{doping}). The trapped state still 
corresponds to a photo-doped state, but now at an effective temperature above $T_N^\text{doped}$.  
In this context is also interesting to look at the timescale $\tau$ for melting of the antiferromagnetic order, which diverges like $\tau^{-1} 
\propto U-U_c$ close to the transition (bottom panel of Fig.~\ref{doping}). This divergence resembles the critical slowing down at the equilibrium phase transition,
which is related to the divergence of correlation times in equilibrium.

\subsection{Implications for superconducting states}

Finally, let us comment on implications of these results for other symmetry broken states, in particlular $s$-wave superconductivity (SC). 
As mentioned in the introduction, the AFM phase of the half-filled repulsive Hubbard model can be mapped onto the $s$-wave SC phase 
in the half-filled attractive model. The mapping is a particle-hole transformation for the up-spin on the bipartite lattice: 
$c_{i\uparrow}\rightarrow \tilde c^\dagger_{i\uparrow}$ for $i\in A$, $c_{i\uparrow}\rightarrow -\tilde c^\dagger_{i\uparrow}$ for $i\in B$, 
where $A$ and $B$ are sublattice indices.\cite{Shiba1972} Applying this transformation to the repulsive model yields an attractive Hubbard model with 
$\tilde U=-U$. A photo-doped antiferromagnet with doublons and holes transforms into a superconducting or charge-ordered state with an enhanced number of unpaired electrons. 
Our finding of long-lived nonthermal AFM states in the repulsive model therefore implies the existence of similar long-lived nonthermal SC states after quenches within the strongly-interacting regime on the attractive side. The similarity of the trapped AFM state to a photo-doped 
Mott insulator in turn suggests that a similar trapped SC state should appear after a photo-induced pair-breakup in the strongly interacting attractive 
regime. The stability of this nonthermal SC is then linked to the slow recombination of unpaired electrons. 

While it is clear that our simple model (with $s$-wave superconductivity) and the single-site DMFT formalism are in many respects not appropriate to describe the 
complicated physics of underdoped cuprates, one can speculate that some of the physics discussed in this paper may be at play in the recently reported experiment by the Cavalleri group,\cite{Fausti2011} where a photo-induced transient SC state in a cuprate material was found to be stable for several tens of pico-seconds, 
and thus much longer than the light-pulse which stimulated apical oxygens and (by a yet unknown mechanism) induced the SC state.
Long-lived, nonthermal symmetry broken states which are 
stabilized by the exponentially long life-time of doublons should also affect the relaxation dynamics in other strongly correlated, long-range ordered systems, such as CDW compounds. 

\acknowledgements

We thank H. Aoki, A. Georges, A. Lichtenstein, J. Mentink, A. J. Millis and T.~Oka for stimulating discussions. The calculations were run on the Brutus cluster at ETH Zurich. We acknowledge support from the Swiss National Science Foundation (Grant PP0022-118866) and FP7/ERC starting grant No. 278023.

\end{document}